\documentclass{iau}
\usepackage{graphicx}
\usepackage{amsmath}

\usepackage{url}

\usepackage{xspace}

\def\apj{ApJ}
\def\apjl{ApJ}
\def\apjs{ApJS}

\def\aap{A\&A}

\def\aaps{A\&AS}

\def\mnras{MNRAS}

\def\nat{Nature}

\def\rmxaa{RMxAA}

\newcommand{\maihem}{\textsc{maihem}\xspace}
\newcommand{\flash}{\textsc{flash}\xspace}
\newcommand{\cloudy}{\textsc{cloudy}\xspace}

\newcommand{\ionic}[2]{#1$\,${\scshape{#2}}\xspace}%    % ion, i.e., C II = \ionic{C}{ii}
\newcommand{\ionf}[2]{#1$\,${\scshape{#2}}}

\usepackage{natbib}
\usepackage{color,hyperref}
\hypersetup{breaklinks=true,colorlinks=true,
            linkcolor=blue,urlcolor=blue,anchorcolor=blue,citecolor=blue,filecolor=blue,
            bookmarks=true, bookmarksopen=true, bookmarksopenlevel=1}  

\title[Hydrodynamic Simulations of Starburst-driven Superwinds] %% give here short title %%
{Hydrodynamic Simulations and Time-dependent Photoionization Modeling of Starburst-driven Superwinds}

\author[A. Danehkar, M. S. Oey, \& W. J. Gray]   %% give here short author list %%
{A. Danehkar, M. S. Oey, and W. J. Gray}

\affiliation{Department of Astronomy, University of Michigan, Ann Arbor, MI 48109, USA
\\email: {\tt \href{mailto:danehkar@eurekasci.com}{danehkar@eurekasci.com}}}

\pubyear{2023}
\volume{362}  %% insert here IAU Symposium No.
\pagerange{64-69}
\doi{\href{https://doi.org/10.1017/S1743921322001570}{10.1017/S1743921322001570}}
\jname{Predictive Power of Computational Astrophysics \\as a Discovery Tool}
\editors{D. Bisikalo, D. Wiebe \& C. Boily, eds.}
\begin{document}

\maketitle

\begin{abstract}
Thermal energies deposited by OB stellar clusters in starburst galaxies lead to the formation of galactic superwinds. Multi-wavelength observations of starburst-driven superwinds pointed at complex thermal and ionization structures which cannot adequately be explained by simple adiabatic assumptions. In this study, we perform hydrodynamic simulations of a fluid model coupled to radiative cooling functions, and generate time-dependent non-equilibrium photoionization models to predict physical conditions and ionization structures of superwinds using the \maihem atomic and cooling package built on the program \flash. Time-dependent ionization states and physical conditions produced by our simulations are used to calculate the emission lines of superwinds for various parameters, which allow us to explore implications of non-equilibrium ionization for starburst regions with potential radiative cooling.
\keywords{Stars: winds, outflows -- galaxies: starburst -- hydrodynamics -- ISM: bubbles -- radiation mechanisms: general -- galaxies: star clusters -- intergalactic medium}
\end{abstract}

\firstsection % if your document starts with a section,
              
\section{Introduction}

Thermal and mechanical feedback from OB stars in stellar clusters displaces 
the surrounding medium in starburst regions on a large scale and forms a galactic-scale outflow named
superwind \citep{Heckman1990}, accompanied by a narrow shell and sometime by a hot bubble called 
superbubble \citep{Weaver1977}.
The physical properties of the expanding wind region prior to the bubble and shell have been obtained by \citet{Chevalier1985} using adiabatic fluid equations that yield the outflow density $n\varpropto r^{-2}$ and 
temperature $T\varpropto r^{-4/3}$. However, the fluid equations coupled to radiative cooling functions studied by \citet{Silich2004} depict a deviation from the adiabatic temperature, which may explain strong cooling and suppressed superwinds seen in observations of some star-forming galaxies \citep{Oey2017,Turner2017,Jaskot2017}. 
In particular, semi-analytic studies and hydrodynamic simulations demonstrated that radiative cooling is heavily dependent on the metallicity, mass-loss rate, and wind velocity \citep{Silich2004,Tenorio-Tagle2005,Gray2019a,Danehkar2021}.
In the case of starburst galaxies where metallicity is low, high mass-loss rates and low outflow velocities 
contribute to substantial radiative cooling \citep{Danehkar2021}.

Photoionization calculations were performed to identify the superwind models with strong radiative cooling \citep{Gray2019a,Danehkar2021}. However, emission lines in photoionization (PIE) and collisional ionization equilibrium (CIE) calculated by \citet{Danehkar2021} did not make a clear distinction between those with and without substantial radiative cooling. Previously, photoionization models built with time-dependent non-equilibrium ionization (NEI) states by \citet{Gray2019a} and \citet{Gray2019} also indicate that ions such as \ionic{O}{vi} and \ionic{C}{iv} could behave differently where plasma is in the NEI case.
Non-equilibrium conditions occur when the radiative cooling timescale $\tau_{\rm cool}$ is shorter than the CIE timescale $\tau_{\rm CIE}$ \citep[see e.g.,][]{Gnat2007}). In the expanding wind region where plasma is in transition from CIE to PIE at temperatures below $10^6$\,K, NEI conditions may emerge \citep{Vasiliev2011}
and NEI states could have substantial deviations from CIE states \citep{Gnat2007,Vasiliev2011,Oppenheimer2013}, which can be identified using the \ionic{C}{iv}, 
\ionic{N}{v}, and \ionic{O}{vi} lines.

Recently, \citet{Danehkar2021} reported emission line fluxes calculated based on CIE$+$PIE assumptions using the physical conditions obtained from hydrodynamic simulations with the \maihem module in the \flash program. Similarly, we also computed emission line fluxes following \citet{Gray2019a} for NEI conditions using the time-dependent NEI states and physical properties predicted by our hydrodynamic simulations \citep{Danehkar2022}, showing that the \ionic{C}{iv} and \ionic{O}{vi} lines have different behaviors in non-equilibrium photoionized expanding wind regions.

\section{Hydrodynamic Simulations}

To study starburst-driven superwinds, we consider starburst feedback from a spherically symmetric stellar cluster parameterized by the cluster radius $R_{\rm sc}$, the mass-loss rate $\dot{M}$, and the stellar wind velocity $V_{\infty}$.
The radiation field is characterized by the total stellar luminosity  and spectral energy distribution (SED).
The surrounding medium has a density $n_{\rm amb}$, while its temperature $T_{\rm amb}$ is dependent on the radiation field and determined by a \cloudy model. 

Our hydrodynamic simulations are performed using 
the \maihem atomic and cooling package \citep{Gray2015,Gray2016,Gray2019} in the frame work of the \flash program \citep{Fryxell2000}, which obtains the solutions for the following one-dimensional spherically symmetric fluid equations coupled to the radiative cooling and photo-heating functions:
\begin{align}
\frac{d \rho}{d t}+\frac{1}{r^2} \frac{d}{dr} \left( \rho u r^2 \right) & = q_{m},  \label{eq_1} \\ 
\frac{d \rho u}{d t}+\rho u \frac{d u}{dr} + \frac{d P}{dr} & = - q_{m} u, \label{eq_2} \\ 
\frac{d {\rho E}}{d t} + \frac{1}{r^2} \frac{d}{dr} \left[ \rho u r^2  \left( \frac{u^{2}}{2} +\frac{\gamma}{\gamma -1} \frac{P}{\rho}  \right) \right] & =\displaystyle\sum_{i}^{}  n_i  \Gamma_i-\displaystyle\sum_{i}^{}  n_i n_e \Lambda_i + q_{e}, \label{eq_3} 
\end{align}
where $r$ is the radius, $\rho$ the density, $u$ the velocity, $P$ the thermal pressure, $E$ the total energy per unit mass, $\gamma=5/3$ the specific heat ratio, $q_{m} = \dot{M} / (\frac{4}{3} \pi R^3_{\rm sc})$ and $q_{e} = ( \frac{1}{2} \dot{M} V_{\infty}^2 ) / (\frac{4}{3} \pi R^3_{\rm sc})$ the mass and energy deposition rate per unit volume, respectively, $n_i$ the number densities of ions,  $n_e$ the electron number density, $\Lambda_i$ the radiative cooling rates for a specified temperature from \citet{Gnat2012}, $\Gamma_i = \int^{\infty}_{\nu_{0,i}} (4 \pi J_{\nu}/h\nu) h (\nu-\nu_{0,i}) \sigma_{i}(\nu) {\rm d}\nu$ 
the photo-heating rates obtained from the given radiation field $J_{\nu}$ and the photoionization cross-section  $\sigma_{i}(\nu)$ \citep{Verner1995,Verner1996}, $\nu$ the frequency, $\nu_{0,i}$ the ionization frequency, and $h$ the Planck constant.

To set the boundary conditions, we employ the semi-analytic radiative assumptions adopted by \citet{Silich2004}, which are based on the adiabatic solutions obtained by \citet{Chevalier1985}. 
Accordingly, the density, temperature, and velocity at $r=R_{\rm sc}$ are set to $\rho =  \dot{M}/ (2 \pi  R_{\rm sc}^2  V_{\infty})$, $T =  (\frac{1}{2} V_{\infty})^2 \mu /(\gamma k_{\rm B})$, and $u = \frac{1}{2} V_{\infty}$, respectively ($\mu$ the mean mass per particle, and $k_{\rm B}$ the Boltzmann constant). For the initial conditions, we set the ambient density specified by an input parameter and the ambient temperature determined by our \cloudy model, while the medium outside the cluster radius is in stationary states ($u=0$) at $t=0$.

\begin{figure}%[b]
\centering
\includegraphics[width=0.45\textwidth, trim = 0 0 0 0, clip, angle=270]{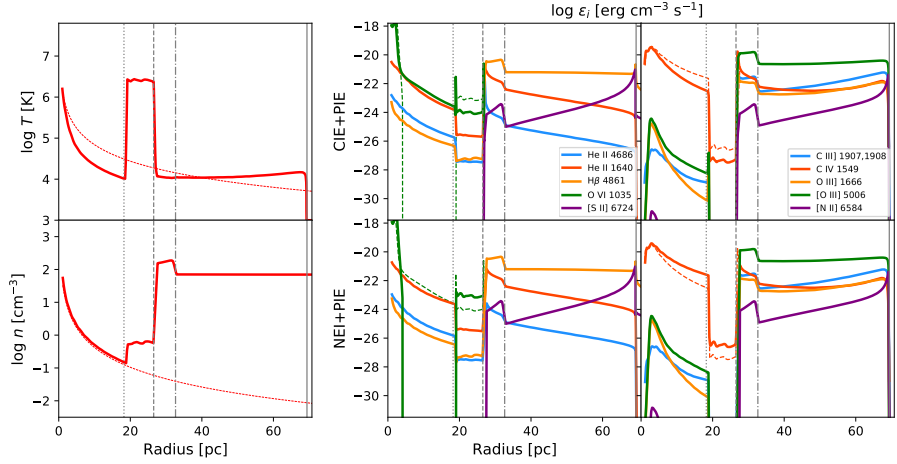}
\caption{\textit{Left Panels}: The temperature $T$ and density $n$ profiles (solid red lines) predicted by our \maihem simulations
 with the adiabatic solutions (red dashed lines). \textit{Right Panels}: The line emissivities $\varepsilon_{i}$ of the emission lines \ionic{He}{ii} $\lambda$4686,$\lambda$1640, H$\beta$ $\lambda$4861, \ionic{O}{vi} $\lambda$1035, [\ionf{S}{ii}] $\lambda$6724 (left), \ionf{C}{iii}] $\lambda\lambda$1907,1908, \ionic{C}{iv} $\lambda\lambda$1549, \ionf{O}{iii}] $\lambda$1666, [\ionf{O}{iii}] $\lambda$5006, and [\ionf{N}{ii}] $\lambda$6584 (right) calculated by the model in collisional ionization and photoionization equilibrium \citep[CIE$+$PIE;][]{Danehkar2021}, and in non-equilibrium photoionization \citep[NEI with PIE in the ambient medium;][]{Danehkar2022}.
The bubble boundaries, the shell end, and Str\"{o}mgren sphere are depicted by dotted, dashed, dash-dotted, and solid vertical gray lines, respectively. 
The wind parameters are: $V_{\infty}=457$\,km\,s$^{-1}$, $\dot{M}= 0.607 \times 10^{-2}$\,M$_{\odot}$\,yr$^{-1}$, $t=1$\,Myr, the stellar cluster $R_{\rm sc} =1$\,pc and $M_{\star}=2 \times 10^6$\,M$_{\odot}$, and the medium  $n_{\rm amb}=100$\,cm$^{-3}$ and $Z/Z_{\odot}=0.5$.
The \ionic{O}{vi} and \ionf{C}{iv} lines predicted by NEI are overplotted by dashed lines in the CIE+PIE panel, and vice versa.
}
\label{fig1}
\end{figure}

The radiation field $J_{\nu}$ included in our hydrodynamic simulations is generated by the stellar population synthesis program Starburst99 \citep{Levesque2012,Leitherer2014} for the rotational stellar models \citep{Ekstroem2012,Georgy2012} with an initial mass function with slope $\alpha= 2.35$ within the range 0.5--150 M$_{\odot}$ and the total stellar mass $M_{\star}=2\times10^6$\,M$_{\odot}$, which are associated with the mass-loss rate $\dot{M} = 10^{-2}$\,M$_{\odot}$\,yr$^{-1}$ at 1 Myr.
The Starburst99 radiation field is applied to the photo-heating rates in \maihem to perform non-equilibrium calculations, as well as \cloudy photoionization models.

Figure~\ref{fig1} (left panels) presents the temperature $T$ and density $n$ radial profiles (solid red lines)
generated by our \maihem simulation for a model with substantial radiative cooling, along with the expected adiabatic solutions without radiative cooling (dashed lines). The four distinctive regions of a typical superwind defined by \citet{Weaver1977} are also separated by dotted, dashed, and dash-dotted vertical gray color lines, namely expanding wind region (before dotted), bubble (between dotted and dashed), shell (between dashed and dash-dotted), and ambient medium (after dash-dotted vertical lines). 
The Str\"{o}mgren sphere (solid vertical gray color line) are determined by a \cloudy photoionization run on the density profile (pure PIE) following \citet{Danehkar2021}.

\begin{figure}
\centering
\includegraphics[width=0.45\textwidth, trim = 0 0 0 0, clip, angle=270]{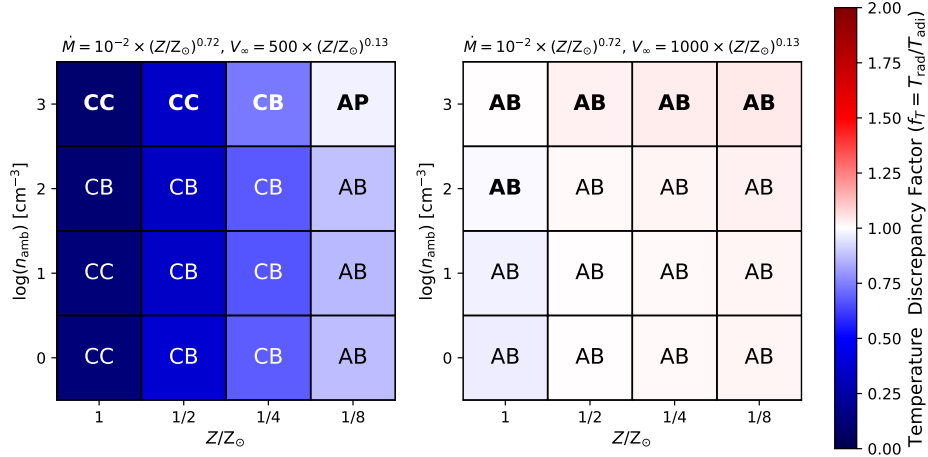}
\caption{The mean radiative temperature $T_{\rm rad}$ with respect to the mean adiabatic temperature $T_{\rm adi}$ of the expanding wind region for various wind parameters $V_{\infty}=500$, and 1000 km\,s$^{-1}$ and  $\dot{M}= 10^{-2} \times (Z/$Z$_{\odot})^{0.72} $\,M$_{\odot}$\,yr$^{-1}$, metallicity $Z$/Z$_{\odot}=0.125$, 0.25, 0.5, and 1, and ambient media with $\log n_{\rm amb}=0$, 1, 2, and 3 \,cm$^{-3}$, surrounding a stellar cluster characterized by $R_{\rm sc} =1$\,pc, $M_{\star}=2 \times 10^6$\,M$_{\odot}$, and age $t=1$ Myr.
The wind models are identified as adiabatic bubble (AB), catastrophic cooling (CC),  catastrophic cooling bubble (CB), and pressure-confined (AP), with optically-thick status (bold font), based on criteria defined by \citet{Danehkar2021}. 
}
\label{fig2}
\end{figure}

Figure~\ref{fig2} shows the mean radiative temperature over the mean adiabatic temperature, $f_T \equiv T_{\rm wind} / T_{\rm adi}$,
of the expanding wind region predicted by our \maihem hydrodynamic simulations for different wind parameters ($V_{\infty}$ and  $\dot{M}$), ambient densities ($n_{\rm amb}$), metallicity ($Z$/Z$_{\odot}$), a stellar cluster with $R_{\rm sc} =1$\,pc and $M_{\star}=2 \times 10^6$\,M$_{\odot}$, and current age $t=1$ Myr.
The catastrophic cooling (CC) and catastrophic cooling bubble (CB) wind modes, which are with and without bubbles, have $f_T< 0.75$, while the adiabatic bubble (AB) and pressure-confined (AP) mode have $0.75<f_T <1.25$. Moreover, the adiabatic pressure-confined (AP) mode is assigned to those models where the bubble expansion is confined by the ambient thermal pressure \citep[for more detail see][]{Danehkar2021}. Optically-thick models having neutral ambient medium are also displayed with the bold font. 
It can be seen that increasing mass-loss rates and decreasing wind velocities result in enhanced radiative cooling in the expanding wind region.

\section{Time-dependent Photoionization Models}

The physical conditions and NEI states of the expanding wind region made by our hydrodynamic simulations with \maihem, along with the radiation field generated by Starburst99, are used to construct non-equilibrium photoionization models with 
the \cloudy program \citep{Ferland2013,Ferland2017}.
Excluding the NEI states, \citet{Danehkar2021} incorporated the physical properties into a grid of \cloudy models, which describe the CIE$+$PIE cases. Inclusion of the NEI states predicted by \maihem allows us to emulate non-equilibrium
photoionization in the expanding wind region \citep{Danehkar2022} for which pure PIE is still present in the ambient medium.

In the non-equilibrium conditions, gas kinematic and ionization structures are closely interconnected. The time-dependent NEI states are linked to collisional and dielectronic recombination, collisional ionization, and photoionization rates.  The number density of ions $n_{i}$ for each chemical element in the non-equilibrium cases can be described as 
\begin{equation}
\frac{1}{n_{\rm e}}\frac{d n_{i}}{d t} = n_{i+1}\alpha_{i+1} - n_{i}\alpha_{i} + n_{i-1} S_{i-1} - n_{i}S_{i} + \frac{1}{n_{\rm e}} n_{i-1} \zeta_{i-1} - \frac{1}{n_{\rm e}} n_{i} \zeta_{i} ,  \label{eq_4}
\end{equation}%
where $\alpha_{i}$ includes the radiative recombination rate \citep{Badnell2006} and dielectronic recombination rates  \citep[references given in Table 1 of][]{Gray2015} for the ionic species ${i}$, $S_{i}$ is the collisional ionization rates \citep{Voronov1997}, $\zeta_{i}= \int^{\infty}_{\nu_{0,i}} (4 \pi J_{\nu}/h\nu) \sigma_{i}(\nu) {\rm d}\nu$ is the photoionization rates calculated using the specified background radiation field $J_{\nu}$ made by Starburst99 and the photoionization cross-section $\sigma_{i}(\nu)$.

The non-equilibrium cases appear when the CIE timescale $\tau_{\rm CIE} \approx 1 / (n_{\rm e}\alpha_{i} + n_{\rm e}S_{i})$ \citep{Mewe1999} is longer than the cooling timescale $\tau_{\rm cool} = 3 (n_i + n_{\rm e}) k_{\rm B} T/ (2 n_i^{2} \Lambda_i)$ \citep{Dopita2003}. For less dense environment ($\lesssim 1$\,cm$^{-3}$) that is typical of the expanding wind region, the ions \ionic{C}{iv}, \ionic{N}{v}, and \ionic{O}{vi} satisfy the condition $\tau_{\rm CIE}\geq \tau_{\rm cool} $ at temperatures below $10^{6}$\,K, so they could be in the NEI situations in the presence of strong radiative cooling. 
For $\tau_{\rm CIE} \ll \tau_{\rm cool} $, plasma is in the CIE conditions. % ($\partial n_{i}/\partial t \approx 0$).

Figure~\ref{fig1} (right panels) presents the emissivities of H$\beta$, low-excitation [\ionf{S}{ii}] and [\ionf{N}{ii}], and high-excitation \ionic{He}{ii}, [\ionf{O}{iii}], and \ionf{C}{iii}], as well as 
highly-ionized \ionic{C}{iv} and \ionic{O}{vi} calculated by \cloudy for the physical properties (without the NEI states) produced by our \maihem simulation associated with plasma in photoionization and collisional ionization equilibrium (top panel; CIE$+$PIE), as well as the emissivities computed using \cloudy following the method of \citet{Gray2019a} for the physical conditions and NEI states generated by our \maihem simulation 
corresponding to the non-equilibrium photoionization situations (bottom panel; NEI with pure PIE in the ambient medium). Large grids for various model parameters are provided as interactive figures by \citet{Danehkar2021} for combined CIE$+$PIE conditions and \citet{Danehkar2022} for combined NEI$+$PIE situations, and hosted on this website \url{https://galacticwinds.github.io/superwinds/}. 
As seen in Figure~\ref{fig1}, the \ionic{O}{vi} and \ionic{C}{iv} emissivity profiles in NEI are not the same as those with CIE (see green and orange dashed lines), particularly in the expanding wind region affected by radiative cooling. This behavior can be explained by 
the time-dependent ionization states of the ions \ionic{O}{vi} and \ionic{C}{iv} at temperatures below $10^{6}$\,K when rapid cooling faster than ionization processes occurs ($\tau_{\rm cool}  < \tau_{\rm CIE}$).

\section{Implications for Starburst Galaxies}

Our NEI calculations indicate that radiative cooling could enhance the \ionic{C}{iv} 1550\,{\AA}  
in metal-rich and the \ionic{O}{vi} 1035\,{\AA}  doublet in metal-poor environments.
The \ionic{C}{iv} emission line were found in some metal-poor starburst galaxies that are good candidates for suppressed or minimal wind signatures \citep{Senchyna2017,Berg2019a,Berg2019b}.
As discussed by \citet{Gray2019a}, these observations might be associated with kinematic features of suppressed bipolar superwinds rather than resonant scattering mentioned by \citet{Berg2019a}.
An \ionic{O}{vi} absorption line associated with a weak outflow was found in Haro\,11, while the \ionic{O}{vi} emission luminosity suggests some cooling loss \citep{Grimes2007}.
The \ionic{O}{vi} $\lambda$1035 doublet absorption identified in a gravitationally lensed, 
galaxy has the features of weak low-ionization winds \citep{Chisholm2018},
which may also be explained by the mass-loss effect under non-equilibrium conditions.
Moreover, observations of a star-forming galaxy depict an extended halo in the \ionic{O}{vi} image, and 
a weak \ionic{O}{vi} absorbing outflow in the spectrum \citep{Hayes2016}, 
which might be an indication of suppressed winds.

The enhancements of the \ionf{C}{iv} and \ionic{O}{vi} lines could be related to substantial radiative cooling
as suggested by \citet{Gray2019a} and \citet{Gray2019}.
Time-dependent NEI states calculated by \citet{deAvillez2012} 
also imply that \ionic{O}{vi} can be produced in NEI at $10^{4.2-5}$\,K below the temperatures that produce \ionic{O}{vi}  in CIE.
Similarly, our NEI calculations (before the shell; see Figure~\ref{fig1}) also show that the \ionic{C}{iv} and \ionic{O}{vi} emission lines do not behave the same in CIE and NEI, especially in the outflow region strongly impacted by radiative cooling.

Time-dependent ionization processes could also be sensitive to time-evolving ionizing sources. Our radiation field was made by  Starburst99 for a typical age of 1\,Myr. The radiation field calculated by Starburst99 can evolve with the age, which can affect the formation of radiative cooling in starburst-driven superwinds. Our future hydrodynamic simulations with time-evolving radiation fields will help us to understand better the implication of time-dependent non-equilibrium ionization for star-forming regions.

%\section*{Acknowledgments}

%\bibliographystyle{mn2e}
%\bibliography{references}

\end{document}